\pdfoutput=1 
\documentclass[aps,showpacs,showkeys,twocolumn,amsmath,amssymb]{revtex4-1}
\usepackage{latexsym}
\usepackage[T2A]{fontenc}
\usepackage[koi8-r]{inputenc}
\usepackage{graphicx}
\usepackage{epsfig}
\usepackage{textcomp}
\usepackage{amssymb}
\usepackage{amsmath}
\usepackage{mathrsfs}
\usepackage{longtable}

\newcommand{\dd}{\mathrm{d}}
\newcommand{\dv}[2]{\frac{\dd#1}{\dd#2}}

\newcommand{\fig}{fig.~}

\newcommand{\Joule}{\text{J}}

\newcommand{\mV}{\text{mV}}

\newcommand{\uA}{\mu\text{A}}

\newcommand{\nW}{\text{nW}}

\newcommand{\Ohm}{\Omega}

\newcommand{\mOhm}{\text{m}\Ohm}

\newcommand{\uH}{\mu\text{H}}

\newcommand{\nF}{\text{nF}}

\newcommand{\s}{\text{s}}
\newcommand{\ps}{\text{ps}}

\newcommand{\um}{\mu\text{m}}
\newcommand{\nm}{\text{nm}}

\newcommand{\cm}{\text{cm}}
\newcommand{\GHz}{\text{GHz}}

\newcommand{\K}{\text{K}}

\newcommand{\dB}{\text{dB}}

\newcommand{\Ic}{I_\text{c}} 
\newcommand{\Tc}{T_\text{c}} 
\newcommand{\Te}{\Theta} 
\newcommand{\Tp}{T_\text{p}} 
\newcommand{\Tpt}{T^\text{p}} 
\newcommand{\Tb}{T_\text{b}} 
\newcommand{\Ce}{C_\text{e}} 
\newcommand{\Cp}{C_\text{p}} 
\newcommand{\tep}{\tau_\text{eph}} 

\renewcommand{\fig}{FIG.\,}
\newcommand{\Ubias}{U_{\text{bias}}}
\newcommand{\IL}{I^{\text{L}}}
\newcommand{\ILb}{I_{\text{L}}}

\newcommand{\Iheb}{I^{\text{HEB}}}
\newcommand{\Ihebb}{I_{\text{HEB}}}

\newcommand{\Rheb}{R^{\text{HEB}}}
\newcommand{\Rhebb}{R_{\text{HEB}}}

\newcommand{\Rhebbc}{\breve{R}_{\text{HEB}}}
\newcommand{\Ild}{I^{\text{load}}}
\newcommand{\Ildb}{I_{\text{load}}}
\newcommand{\Rld}{R^{\text{load}}}
\newcommand{\Rldb}{R_{\text{load}}}

\newcommand{\Ps}{P_{\text{s}}}
\newcommand{\Psloc}{\breve{P}_{\text{sLO}}}
\newcommand{\Plo}{P_{\text{LO}}}
\newcommand{\Pslo}{P_{\text{sLO}}}

\newcommand{\Pldb}{P_{\text{load}}}

\newcommand{\Cdc}{C_{\text{dc}}}
\newcommand{\Crf}{C_{\text{rf}}}

\newcommand{\Crfc}{\breve\Crf}

\begin{document}
\title{RF heating efficiency of the terahertz superconducting hot-electron bolometer}

\author{Sergey~Maslennikov}
\email[]{onduty@rplab.ru}
\affiliation{Moscow State Pedagogical University~(MSPU), ulitsa Malaya Pirogovskaya, dom 29, Moskva, 119435, Russia}

\begin{abstract}
We report results of the numerical solution by the Euler method of the system of heat balance equations written in recurrent form for the superconducting hot-electron bolometer (HEB) embedded in an electrical circuit. By taking into account the dependence of the HEB resistance on the transport current we have been able to calculate rigorously the RF heating efficiency, absorbed local oscillator (LO) power and conversion gain of the HEB mixer. We show that the calculated conversion gain is in excellent agreement with the experimental results, and that the substitution of the calculated RF heating efficiency and absorbed LO power into the expressions for the conversion gain and noise temperature given by the analytical small-signal model of the HEB yields excellent agreement with the corresponding measured values.
\end{abstract}

\pacs{not assigned}

\keywords{superconducting hot-electron bolometer mixer, HEB, NbN, distributed model, heat balance equations, conversion gain, RF heating efficiency, noise temperature, simulation, Euler method}

\maketitle{}

Investigations of terahertz superconducting hot-electron bolometers (HEB)~\cite{first_heb} are motivated by astrophysics, in particular, by the fact that a half the radiation and $98\,\%$ of the photons coming towards the Earth lie in the terahertz range~\cite{motivation_interfrm}.
HEBs have proved to be perfect mixers for terahertz astronomy and heterodyne spectroscopy because of their excellent noise performance and low local oscillator (LO) power requirement~\cite{DrakinskiyCherednichenkoY2008id908,HoninghPuetzY2012id907}.
At the same time, an adequate understanding of the underlying physical mechanisms and a reasonable description of the HEB mixer conversion gain and noise temperature have not been reached yet~\cite{MerkelKhosropanahY2000id893,Khosropanah_thes}. The key to explain these main characteristics of the HEB is its RF heating efficiency~\cite{MerkelKhosropanahY2000id893,Khosropanah_thes}.

Typically, an HEB is a rectangular bridge made of a thin superconducting film deposited on a silicon substrate and integrated with a normal-metal planar antenna~\cite{RyabchunTretyakovY2011id716,TretyakovRyabchunY2009id599}.
The HEB is modeled by resistivity, electronic specific heat, thermal conductivity, temperature, and other important physical quantities distributed along the bridge~\cite{MerkelKhosropanahY2000id893,Khosropanah_thes}. The behaviour of the HEB mixer at the intermediate frequency (IF) is described by the ``distributed model'' (DM)~\cite{MerkelKhosropanahY2000id893,Khosropanah_thes}. It is argued that the DM is inconsistent~\cite{MerkelKhosropanahY2000id893,Khosropanah_thes} i.e. for the bolometer resistance $\Rhebb$ and the absorbed RF power $\Pslo$ substitution of the calculated RF heating efficiency
\begin{equation}\label{eq_Crf}
 \Crf = \dv{\Rhebb}{\Pslo}
\end{equation}
into the analytical expressions for the conversion gain and noise temperature~\cite{MerkelKhosropanahY2000id893,Khosropanah_thes} produces results inconsistent with the experiment~\cite{MerkelKhosropanahY2000id893,Khosropanah_thes}. Although the RF heating efficiency of the HEB is the key parameter,
there is no an appropriate calculation approach within the known models~\cite{Khosropanah_thes} that would predict it.
It is shown that to make the DM capable of predicting the measured dependencies of the HEB conversion gain and noise temperature the value of $\Crf$ has to be at least 3 times smaller than that given by the DM~\cite{MerkelKhosropanahY2000id893,Khosropanah_thes}. This problem has not been resolved till now.
Below we show how one can calculate $\Crf$ using the DM.

\begin{figure}[!b]
\begin{center}
 \includegraphics[width=0.9\columnwidth]{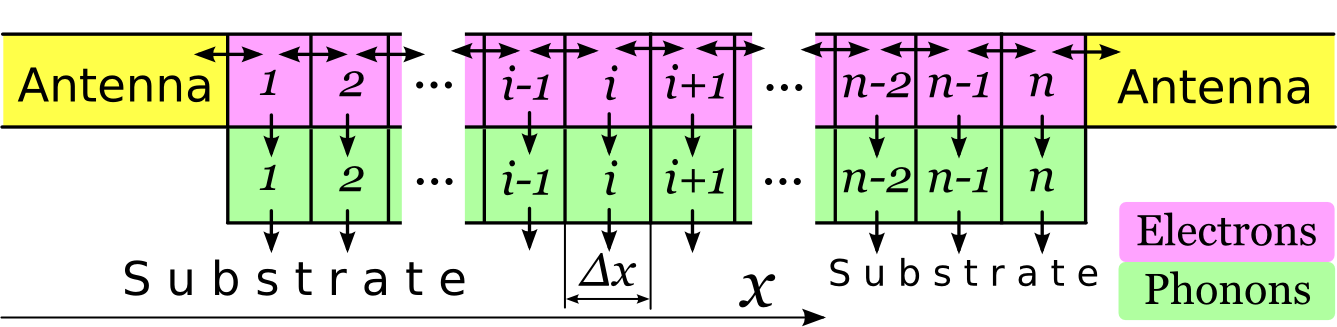}
\end{center}
\caption{\label{fig_blnc}Heat balance in the superconducting HEB. The arrows denote heat flows for diffusion (between elementary cells of electron subsystem), for electron-phonon cooling (from electrons to phonons), and for non equilibrium phonons escape to the substrate (from phonons to substrate).}
\end{figure}

We simulate the same superconducting HEB as was simulated in the work of P.\,Khosropanah~\cite{MerkelKhosropanahY2000id893,Khosropanah_thes} where the HEB is considered as a NbN strip integrated with a planar antenna~\cite{RyabchunTretyakovY2011id716,ifbandwidth_6G,hot_electrons_Semenov}. Since the diffusion is one of the cooling channels~\cite{HEB_diffusion_NbC,MerkelKhosropanahY2000id893,Khosropanah_thes,TretyakovRyabchunY2009id599} of the superconducting HEB, one does not consider it as a lumped element but splits it in the model into the cells in order to consider the heat flow in each cell separately~(\fig\ref{fig_blnc}).
In this approach the energy is absorbed by electrons of the elementary spatial cell and after thermalization of the electrons~\cite{LindgrenIlin1Y2000id856} the heat flows to the neighboring cells via diffusion, and to the phonons of the same spatial cell via the electron-phonon interaction~\cite{GershenzonsY1990id902}. The power influx into the cell is also possible from the neighboring cells and from the Joule heating. A similar approach has already been considered~\cite{MerkelKhosropanahY2000id893,Khosropanah_thes,HajeniusBarendsY2005id604} but for the case of the detailed balance only. For the phonon subsystem of the HEB the power influx comes from the electrons, while the power outflow is possible to the substrate, which plays the role of a thermostat.
The DM is governed by the system of recurrent heat-balance equations similar to that for the average-temperature models~\cite{sc_resp_Perrin_1983,hot_electrons_KE_proc_95,SemenovNebosisY1996id605} but written for each elementary cell $i$~(\fig\ref{fig_blnc}) and time $t_j$ in recurrent form:
\begin{equation}\label{eq_blnc}
 \begin{array}{ll}
  \Te_{i,j+1} \approx & \Te_{i,j} +\\
  & + \frac{1}{\Ce(\Te_{i,j})}
     \left( {\Ihebb^2}_j R(\Iheb_j, \Te_{i,j}) + \right.\\
  &     + \Pslo(t_{j}) -\omega_1(\Te_{i,j}, \Tpt_{i,j}) +\\
  &     + \frac{1}{2\Delta x}
      \left[
        (\varkappa_{i,j} + \varkappa_{i+1,j})(\Te_{i+1,j}-\Te_{i,j}) -\right.\\
  & \left.\left.  - (\varkappa_{i-1,j} + \varkappa_{i,j})(\Te_{i,j}-\Te_{i-1,j})
      \right]
     \right) \Delta t\\
  \Tpt_{i,j+1} \approx & \Tpt_{i,j} +
    \frac{
    \left(
     \omega_1(\Te_{i,j}, \Tpt_{i,j})
     - \omega_2(\Tpt_{i,j}, \Tb)
    \right)}
    {\Cp(\Tpt_{i,j})}
    \Delta t,
 \end{array}
\end{equation}
where $\Te$ and $\Tp$ are the electronic and phonon temperatures, respectively, $\Tb = 4.2\,\K$ is the temperature of substrate and antenna, $\Ihebb$ is the HEB current, $R$ is the cell resistance,  $\Pslo$ is the net absorbed power of the signal and LO~\cite{HajeniusBaselmansY2005,DrakinskiyCherednichenkoY2008id908}, $\omega_1$ and $\omega_2$ are the functions describing the heat flows from electrons to phonons and from phonons to substrate, respectively~\cite{SemenovNebosisY1996id605}, $\Ce$ and $\Cp$ are the electronic and phonon heat capacities, respectively~\cite{SemenovNebosisY1996id605}, $\varkappa$ is the electronic thermal conductivity per unit length, $\Delta x = x_{i+1} - x_i$ is the cell length, $\Delta t = t_{j+1} - t_j$ is the small time interval (here $i$ indexes position, and $j$ indexes time). The exact expressions for
$\omega_1(\Te, \Tpt)$, $\omega_2(\Tpt, \Tb)$, $\Cp(\Tpt)$ are defined in the work of R.\,Nebosis~\cite{SemenovNebosisY1996id605}. The Bardeen-Cooper-Schrieffer (BCS) theory~\cite{CooperBardeenY1957id901} is used to calculate $\Ce(\Te)$ and $\varkappa(\Te)$ expressions numerically (see, e.g., section 3.6.3 of M.\,Tinkham's book~\cite{TinkhamY1996id896}). Calculation of $R(\Iheb_j, \Te_{i,j})$ is discussed below.

The most important idea of this work is to solve the electrical equations together with the heat-balance equations~(\ref{eq_blnc}) numerically by the Euler method. This approach is simple in formulation that reduces the probability of a mistake. It automatically includes the electro-thermal feedback~\cite{SemenovNebosisY1996id605} and does not require additional suppositions to calculate the DC and RF heating efficiencies~\cite{MerkelKhosropanahY2000id893,Khosropanah_thes}.
The electrical equations are deduced from the same electrical circuit diagram of a practical HEB-based receiver~(\fig\ref{fig_diag}) as explained by P.\,Khosropanah and co-authors~\cite{MerkelKhosropanahY2000id893,Khosropanah_thes}:
\begin{figure}[!b]
\begin{center}
\includegraphics[width=0.5\columnwidth]{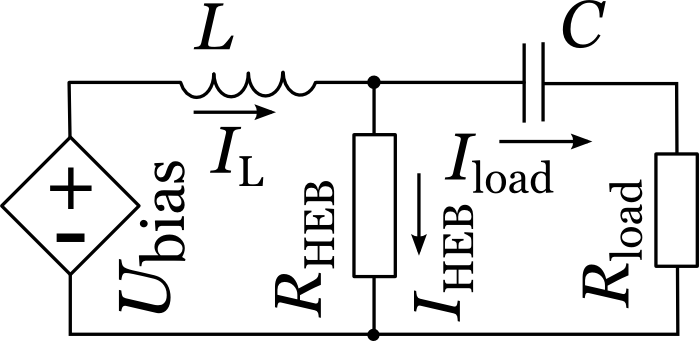}
\end{center}
\caption{\label{fig_diag}An equivalent circuit for the simulation of the superconducting HEB as a direct detector or a mixer in a receiver.}
\end{figure}
\begin{equation}\label{eq_el_1}
 \IL_{j+1} = \IL_j + \frac{1}{L} (\Ubias - \Iheb_j \Rheb_j) \Delta t,
\end{equation}
\begin{equation}\label{eq_el_2}
 Q_{j+1} = Q_j + \Ild_j \Delta t
\end{equation}
\begin{equation}\label{eq_el_3}
 \Ild_{j+1} = \frac{-\frac{1}{C}Q_{j+1} + \IL_{j+1} \Rheb_{j+1}}
                   {\Rld + \Rheb_{j+1}}
\end{equation}
\begin{equation}\label{eq_el_4}
 \Iheb_{j+1} = \IL_{j+1} - \Ild_{j+1}
\end{equation}
In these equations $\Rldb=50\,\Ohm$ is the impedance of the IF-amplifier which plays the role of a matched load in the circuit, $Q$ is the charge of the capacitor C~(\fig\ref{fig_diag}) with capacitance $C=1\,\nF$, $\Ubias$ is the bias voltage, $\ILb$, $\Ildb$, and $\Ihebb$ are the currents of the inductor, load, and HEB, respectively, $L=1\,\uH$ is the inductance of the inductor,
\begin{equation}\label{eq_R_as_sum}
 \Rheb_j = \sum_i R(\Iheb_j, \Te_{i,j}),
\end{equation}
$R(\Iheb_j, \Te_{i,j})$ is the resistance of the elementary cell $i$.

In order to simulate such a circuit one needs the dependence $\Rhebb(\Te)$ to be theoretically predicted or measured for an HEB under the condition of a small current $\Ihebb\ll\Ic$~($\Ic$ is the critical current) and uniform $\Te$. If $\Te$ is uniform one can put $R(\Te) = \Rhebb(\Te) / n$ for each of $n$ equal elementary cells.
However, in the optimal operating condition, the HEB is strongly biased, $\Ihebb$ is not small enough and one has to take into account the dependence of the resistance of an elementary HEB cell on the current~\cite{HajeniusBarendsY2005id604}. A theoretical description of this dependence is complicated by material properties, pinning sites, inhomogeneities, granularity, and finite-size effects, to name just a few, but this dependence can be described empirically by the negative temperature shift~\cite{HajeniusBarendsY2005id604}
\begin{equation}\label{eq_Tc_shift}
 \Delta\Te_j = \left( \frac{\Iheb_j}{\Ic} \right)^{\frac{1}{\gamma}} \Tc
\end{equation}
written for the critical current $\Ic$, critical temperature $\Tc$, and the empirical constant $\gamma=0.54$~\cite{HajeniusBarendsY2005id604}.
In other words, one has to replace the temperature value $\Te_k$ in each point $k$ of the dependence $\Rhebb(\Te_k)$ by $\Te_k - \Delta\Te_j$ where $\Delta\Te_j$ is calculated for each time step $j$.

The experimental basis of this work includes the dependence $\Rhebb(\Te)$~\cite{MerkelKhosropanahY2000id893,Khosropanah_thes} and NbN HEB dimensions~\cite{MerkelKhosropanahY2000id893,Khosropanah_thes}: a length of $0.4\,\um$, a width of $4\,\um$, and a thickness of $5\,\nm$. $\Ce$ is calculated with the NbN electronic diffusion constant~\cite{GershenzonsY1990id902,NbN_heb_det_Gousev_1994,HEB_diffusion_NbC} of $0.45\,\cm^2\,s^{-1}$~\cite{SemenovY2001_Q_detection}. The other properties of NbN included into the model are
the electron-phonon interaction time $\tep = 480\,\ps\,\K^{1.6}\,\Te^{-1.6}$~\cite{NbN_heb_det_Gousev_1994}, and the phonon heat capacity $\Cp=9.8\times10^{-6}\,\Joule\,\cm^{-3}\K^{-4}\,V\,\Tp^3$ where $V$ is the volume of the elementary cell~\cite{SemenovY1995id903}.

For the $\Crf$ calculation the HEB model is driven by absorbed power oscillating at frequency $f$ given by the expression~\cite{MerkelKhosropanahY2000id893,Khosropanah_thes}
\begin{equation}\label{eq_Pslo}
 \Pslo(t) = \Ps + \Plo + 2\sqrt{\Ps\Plo} \cos(2 \pi f t + \varphi_0),
\end{equation}
where $\Ps$ and $\Plo$ are the signal and local oscillator powers, respectively.
The $\Ubias$ variable is put equal to the experimental optimum value of $0.8\,\mV$~\cite{MerkelKhosropanahY2000id893,Khosropanah_thes}, and $\overline\Ihebb$ is set to be close to the optimum value of $40\,\uA$~\cite{MerkelKhosropanahY2000id893,Khosropanah_thes} by adjustment of the $\Plo$ value.

\begin{figure}[!b]
\begin{center}
\includegraphics[width=0.9\columnwidth]{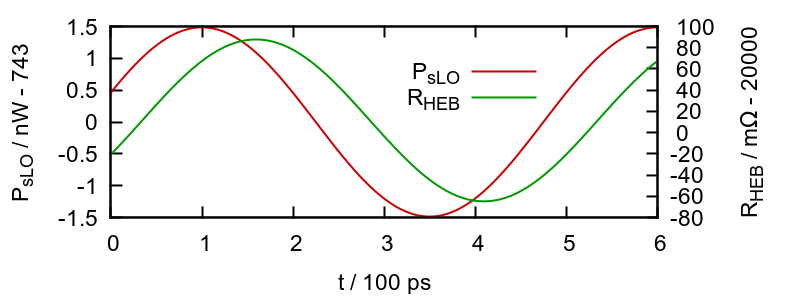}
\end{center}
\caption{\label{fig_resp_tPU}Calculated HEB absorbed power (impact) and resistance (response) dependencies on time $t$.}
\end{figure}

\begin{figure}[!b]
\begin{center}
\includegraphics[width=0.9\columnwidth]{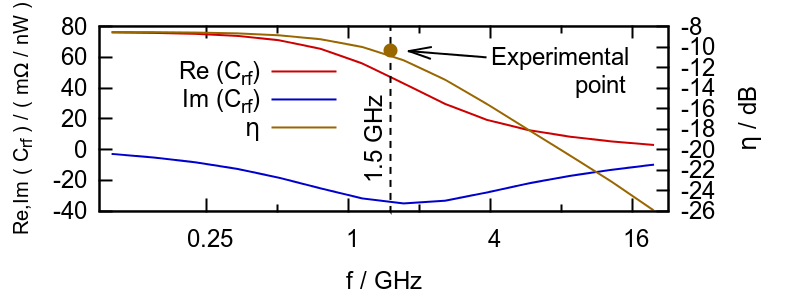}
\end{center}
\caption{\label{fig_freq_Crf_gain}Calculated HEB RF heating efficiency $\Crfc$, conversion gain $\eta$ versus frequency $f$ and an experimental point corresponding to the measured value of the conversion gain at the optimum bias voltage of $0.8\,\mV$~\cite{MerkelKhosropanahY2000id893,Khosropanah_thes}. The experimental point is attributed to the output YIG filter~\cite{Khosropanah_thes} frequency of $1.5\,\GHz$.}
\end{figure}

The system of equations (\ref{eq_blnc}),(\ref{eq_el_1}),(\ref{eq_el_2}),(\ref{eq_el_3}),(\ref{eq_el_4}) is solved by the Euler method. For $f=2\,\GHz$ the solution (response) $\Rhebb(t)$ is shown in \fig\ref{fig_resp_tPU} along with the impact $\Pslo(t)$.
As the impact and response at a frequency $f$ are harmonic functions, the RF heating efficiency of the HEB at this frequency can be calculated as
\begin{equation}
 \Crfc = \Rhebbc / \Psloc,
\end{equation}
where~(from \fig\ref{fig_resp_tPU}) $\Psloc \approx 1.5\,\nW\times e^{i\varphi}$, and $\Rhebbc \approx 76\,\mOhm\times e^{i(\varphi-0.24\,\pi)}$ are complex amplitudes of the HEB absorbed power and resistance, respectively. The RF heating efficiency of the HEB at $f=2\,\GHz$ is close to $51\,\mOhm/\nW\times e^{i(-0.24\,\pi)}$.
Solving the system of equations (\ref{eq_blnc}),(\ref{eq_el_1}),(\ref{eq_el_2}),(\ref{eq_el_3}),(\ref{eq_el_4}) for other values of $f$ and applying the same reasoning one builds the frequency dependence of RF heating efficiency~(\fig\ref{fig_freq_Crf_gain}).

By averaging $\Ildb^2$ it is also easy enough to extract conversion gain from the model. The average power dissipated in the load is $\overline\Pldb=\Rldb\overline{\Ildb^2}$ and the HEB mixer conversion gain is
\begin{equation}\label{eq_gain}
 \eta = \overline\Pldb / \Ps
\end{equation}
Applying the same procedure for different values of $f$, one builds the frequency dependence of calculated conversion gain~(\fig\ref{fig_freq_Crf_gain}).



In the experimental work of P.\,Khosropanah~\cite{MerkelKhosropanahY2000id893,Khosropanah_thes} the noise temperature and conversion gain of the HEB were measured at the IF
of $1.5\,\GHz$ adjusted by the output YIG filter~\cite{Khosropanah_thes}.
At this IF the calculated conversion gain is $-10.7\,\dB$~(\fig\ref{fig_freq_Crf_gain}) while the measured value of the conversion gain at bias voltage of $0.8\,\mV$ (optimum operating point) amounts to $-10.4\,\dB$~\cite{MerkelKhosropanahY2000id893,Khosropanah_thes}. The agreement with the experiment is excellent.


The key expression to investigate the validity of the HEB mixer small signal model is that for the conversion gain~\cite{hot_spot_Chalmers_98,MerkelKhosropanahY2000id893,Khosropanah_thes}:
\begin{equation}\label{eq_gain_a}
 \eta_\text{a} = \frac{
               2{\overline\Ihebb}^2\Rldb\Crf^2\Plo \times
               \frac{\eta_\text{c}(1.5\,\GHz)}
               {\left.\eta_\text{c}\right|_{f\rightarrow 0}}
             }
             {
                \left(\overline\Rhebb+\Rldb\right)^2
                \left(
                   1- \Cdc{\overline\Ihebb}^2\frac{\Rldb - \overline\Rhebb}{\Rldb + \overline\Rhebb}
                \right)^2
             },
\end{equation}
where $\Cdc\approx320\,\mOhm/\nW$~\cite{MerkelKhosropanahY2000id893,Khosropanah_thes} is the heating efficiency for the HEB mixer direct current, and $\eta_\text{c}$ is the value of the conversion gain~(\fig\ref{fig_freq_Crf_gain}) calculated by the Euler method. Substitution of $\Plo=743\,\nW$, $\overline{\Rhebb}=0.8\,\mV/40\,\uA$, and $\left.\Crf\right|_{f\rightarrow0}=76\,\mOhm/\nW$~(\fig\ref{fig_freq_Crf_gain}) calculated by the Euler method into (\ref{eq_gain_a}) yields the value of $\eta_\text{a}\approx-8.6\,\dB$ which is $1.7\,\dB$ higher than the measured value of the conversion gain.


One can conclude that the work of P.\,Khosropanah and his co-authors~\cite{MerkelKhosropanahY2000id893,Khosropanah_thes} does not compromise the distributed model of the superconducting HEB.
The Euler method can be used to calculate the absorbed radiation power, RF heating efficiency and conversion gain of the superconducting HEB.
The method applied in this work is potentially applicable to calculating other main characteristics e.g. the IF impedance or responsivity of the superconducting HEB. In the suggested approach the BCS theory is applied directly with no
empirical approximations
of the electronic thermal conductivity and
specific heat, while other key dependencies predicted theoretically or measured can easily be integrated with the recurrent form of the differential equations controlling the behavior of the HEB.
This work confirms indirectly the NbN electronic diffusion constant of $0.45\,\cm^2/\s$~\cite{SemenovY2001_Q_detection} extracted from the measurements of the second critical magnetic field~\cite{NbN_heb_det_Gousev_1994}, as well as, the measured temperature dependence of the electron-phonon interaction time~\cite{NbN_heb_det_Gousev_1994}. The optimum RF heating efficiency of the NbN HEB mixer with a length of $0.4\,\um$, a width of $4\,\um$, and a thickness of $5\,\nm$ is close to $76\,\mOhm/\nW$ at the low intermediate frequency.

We thank to P.\,Khosropanah for providing the exact value of the frequency of the output YIG filter.
This work is partly supported by: grant of Russian government with contract 14.B25.31.0007 of June 26, 2013; grant of the Russian president НШ-1918.2014.2; grant of NATO EAP. SFPP984068; grant of РФФИ 13-02-91159-ГФЕН\_а.

\bibliographystyle{apsrev-apl}
\bibliography{sn.bib}

\end{document}